\documentclass[twocolumn,superscriptaddress,secnumarabic,amssymb,%
amsmath,nobibnotes,aps,prd,showpacs,nofootinbib]{revtex4}
\usepackage{caption}
\usepackage{subcaption}
\usepackage{graphicx}
\usepackage{bm}
\usepackage{amsmath}
\usepackage{amsfonts}
\usepackage{amssymb}
\usepackage{epstopdf}
\usepackage{color}
\providecommand{\U}[1]{\protect\rule{.1in}{.1in}}
\newcommand{\be}{\begin{equation}}
\newcommand{\ee}{\end{equation}}

\newcommand{\mincir}{\raise
-3.truept\hbox{\rlap{\hbox{$\sim$}}\raise4.truept\hbox{$<$}\ }}
\newcommand{\magcir}{\raise
-3.truept\hbox{\rlap{\hbox{$\sim$}}\raise4.truept\hbox{$>$}\ }}

\newtheorem{remark}{Remark}[section]

\begin{document}

\title{Note on bouncing backgrounds}

\author{Jaume de Haro\footnote{E-mail: jaime.haro@upc.edu}}
\affiliation{Departament de Matem\`atiques, Universitat Polit\`ecnica de Catalunya, Colom 11, 08222 Terrassa, Spain}
\affiliation{Departament de Matem\`atiques, Universitat Polit\`ecnica de Catalunya, Diagonal 647, 08028 Barcelona, Spain}

\author{Supriya Pan\footnote{E-mail: span@research.jdvu.ac.in}}
\affiliation{Department of Mathematics, Raiganj Surendranath Mahavidyalaya, Sudarsharpur, Raiganj, Uttar Dinajpur, West Bengal 733134, India}

\thispagestyle{empty}

\begin{abstract}

The theory of inflation is one of the fundamental and revolutionary developments of modern cosmology that became able to explain many issues of early universe in the context of the standard cosmological model (SCM). However, the initial singularity of the universe, where physics is indefinite, is still obscure in the combined `SCM+inflation' scenario. An alternative to `SCM+inflation' without the initial singularity is thus always welcome, and bouncing cosmology is an attempt of that. The current work is thus motivated to investigate the bouncing solutions in modified gravity theories when the background universe is described by the spatially flat  Friedmann-Lema{\^\i}tre-Robertson-Walker (FLRW) geometry.
We show that the simplest way to obtain the bouncing cosmologies in such spacetime is to consider some kind of Lagrangians whose gravitational sector depends only  on the square of the Hubble parameter of the FLRW universe.
For these modified Lagrangians, the corresponding Friedmann equation, a constraint in the dynamics of the universe, depicts a curve in the phase space $(H,\rho)$, where $H$ is the Hubble parameter and $\rho$ being the energy density of the universe. As a consequence,  a bouncing cosmology is obtained when this curve is closed and crosses  the axis $H = 0$ at least twice, and whose simplest  particular example  is the ellipse  depicting the well-known holonomy corrected Friedmann equation in Loop Quantum Cosmology (LQC). Sometimes, a crucial point in such theories is the appearance of the Ostrogradski instability at the perturbative level, however, fortunately enough, in the present work as long as the linear level of perturbations is concerned, this instability does not appear but such instability may appear at the higher-order of perturbations.   

\end{abstract}


\vspace{0.5cm}


\maketitle

\section{ Introduction}

The hunting for the exact dynamics of the universe is still continuing even after a series of astronomical missions performed during the last several years. The standard cosmological model, probably the simplest universe model yet, cannot explain  the physics of the early universe well. The initial big-bang singularity, the flatness problem, horizon problem, baryon asymmetry, the origin of structure formation of the universe, and several others are seriously related to it. The problem with dark energy related to current accelerating universe also demands a theory beyond the standard cosmological model. It perhaps should be recalled that standard cosmology does not predict the initial singularity, rather we must say that the standard cosmology is incomplete at the extreme stage of the early universe evolution. Precisely, in the framework of standard cosmology, if one continues to go back into the past of the universe, the energy density, temperature of the universe successively increase and become unbounded which results in a state of an infinite energy density, infinite temperature and infinite curvature of the universe, and all physical laws break down $-$ this initial state is known as the big-bang singularity, see \cite{Brandenberger:2011gk} for more details. Thus, one can see that the standard cosmology does not predict the initial singularity, rather, the standard cosmology collapses at big-bang.      
The theory of inflation \cite{guth} was the result of the above contentions  and consequently, it was found that most of the early physics issues can be successfully explained. But however, the initial big-bang singularity is still unanswered.
Clearly, although there is no doubt that inflation might be considered to be an almost successful theory for the early universe, but the initial big-bang singularity 
demands that perhaps we should be open minded for other alternatives to inflation. An alternative approach to the inflationary paradigm \cite{guth} is the 
so-called bouncing cosmologies \cite{nb08, Lehners:2011kr, Brandenberger:2016vhg} where 
the initial big-bang singularity does not appear. {Certainly, 
the concept of such bouncing scenario with  became one of the fascinating topics in modern cosmology which enabled us to travel beyond the standard cosmological model.

 Following this,  a series of investigations were performed in different cosmological frameworks with an aim to see whether the models are capable of allowing such bouncing nature or not.  The list where the viable bouncing scenarios are possible includes several interesting alternative gravitational theories such as $f(T)$ \cite{Hohmann:2017jao}, $f(R)$ \cite{Bamba:2013fha, Nojiri:2017ncd, Odintsov:2015ynk} (One may also see \cite{Pavlovic:2017umo} where a sequence of bounces is also possible in this gravity theory), the unimodular $f(R)$ gravity \cite{Nojiri:2016ygo} and also the Gauss-Bonnet gravity \cite{Oikonomou:2015qha}. In addition to these,  another interesting theory where the cosmological bounce is also allowed is the Einstein-Skyrme-$\Lambda$ model \cite{Ayon-Beato:2015eca}. Thus, it is clear that the bouncing solutions in most of the 
well known gravity theories are allowed. However, 
the simplest example for such bouncing universe 
is achieved from the holonomy corrected loop quantum cosmology (LQC)  \cite{as}, where the Big-Bang singularity is replaced by a nonsingular Big Bounce.

\

The background of a holonomy corrected LQC in a flat FLRW spacetime can be easily mimicked in modified gravity if one works with an invariant scalar that depends only on the square of the Hubble parameter \cite{helling,ds09,haro12} or directly on the Hubble parameter \cite{hap}. In fact, the idea to obtain nonsingular
cosmologies with bounding scalar invariants is the principle of the so-called {\it limiting curvature hypotesis} (see for example one of the original papers \cite{MB} on this subject).
This scalar  could be  the torsion appearing in teleparallelism, where the spacetime is equipped by the 
unusual 
Weitzenb\"ock connection \cite{w} and  a preferred othonormal basis in the tangent bundle of the spacetime manifold must be chosen, or it could be
the extrinsic curvature scalar in the context of 
the Arnowitt-Deser-Misner (ADM) formalism of GR \cite{ADM}. In both formalisms, the scalar reduces to, $-6H^2$, where the background universe is described by the spatially flat FLRW spacetime using the synchronous co-moving coordinates \cite{ha17}.

\

However,  both formulations  essentially suffer from the same problem, namely, they contain preferred coordinate systems.  {It is well established that the teleparallel gravity theory is not invariant under local Lorentz transformations\cite{lsb}. On the other hand, the use} of extrinsic curvature scalar in the ADM formalism of GR requires to fix the slicing \cite{ha17}. Perhaps it might be interesting to refer \cite{Krssak:2015oua, Hohmann:2018rwf} where the authors argue that instead of the pure-tetrad formalism for the $f(T)$-gravity, if in addition, the spin connection is allowed, then the problem with local Lorentz invariance might be resolved. However, we believe that we need to wait for another couple of years for a definite conclusion in order to be sure that no resulting issue is created due to allowing of such nontrivial spin connection. Henceforth, in this work we do not consider the spin 
connection and stick to the usual tetrad formalism of the teleparallel gravity. 
For these reasons, in this note we try to clarify which scalars are really gauge invariants, that means, the scalars which do not depend on the selected coordinate system. From our viewpoint, and following the ideas presented in  \cite{ellis,ellis1}, one finds that this scalar has to be related with the matter component of the universe, which is depicted by the stress tensor. In this way, first of all, we
choose an unitary time-like
vector field,   
as the gradient of the scalar field used in mimetic gravity \cite{mukhanov,mukhanov1}  {(a recent review on mimetic gravity can be found here \cite{Sebastiani:2016ras})} 
 which would be the time-like eigenvector of the stress tensor that always exists for realistic matter  due to the week energy condition \cite{hawking}, and with its covariant derivative, it is possible to build a tensor whose 
quadratic combined contractions lead to our desired scalar, as has been recently done in \cite{langlois}. In fact, to build this scalar,
one could argue that in order to modify the gravitational sector, one has to use the quantities related to the metric tensor rather than the matter. Thus, instead of the time-like eigenvector of the stress tensor, 
 one  could use  the unitary time-like eigenvector of the Ricci tensor,  which   at the level of the background  leads to the same scalar,  although dealing with perturbations  the two different vector fields will give rise to different scalars. The problem of this choice is that this scalar field is dynamic and leads to an invariant depending on the second order derivatives of the field which implies that at the perturbative level, the Ostrogradski's instability may appear. Fortunately, in the present work as long as  the linear level of the perturbations is concerned, the Ostrogradski instability does not appear, however, at the higher-order of perturbations, the Ostrogradski instability may appear.

\

Now, once this scalar is obtained, the procedure to obtain the bouncing backgrounds becomes straight. The simplest way to build the bouncing backgrounds is to consider the closed curves in the phase space $(H,\rho)$ crossing  the axis $H=0$, at least twice; here $H$ and $\rho$ are respectively the Hubble parameter and the energy density of the FLRW universe.
And then for each curve, the integration of  the corresponding first order differential equation given by its modified Lagrangian equation, one gets the analogous reconstructed Lagrangian that effectively describe the  bouncing background.
The structure of the work has the following organization. The mathematical formulation of the bouncing universe resulting from the modified gravity at the background level has been presented in section \ref{sec-MG}. After that in section \ref{sec-bouncing} we discuss the simplest bouncing scenario. Finally, we close the work in section \ref{sec-discuss} with a brief summary.
We note that the units used throughout the paper are $\hbar=c=1$, and $M_{pl}=\frac{1}{\sqrt{8\pi G}}$ is the reduced Planck's mass.

\section{Modified gravity at the background level}
\label{sec-MG}

The equations of General Relativity (GR) can be easily derived  from the variation of the so-called Einstein-Hilbert (E-H)  action 
 \begin{equation}\label{EH-action}
  S=\int\sqrt{-g}\left(\frac{M_{pl}^2}{2}R+{\mathcal L}_{matt}\right) d^4x,
 \end{equation}
 where  $R$ is the scalar curvature, and we have assumed that the matter sector of the universe is described by a scalar field $\phi$ with potential $V(\phi)$ which is minimally coupled to gravity whose Lagrangian is,
 \begin{eqnarray}
  {\mathcal L}_{matt}= \left(-\frac{\phi_{\mu}\phi^{\mu}}{2}-V(\phi)\right).
  \end{eqnarray}

The main idea to modify GR is to obtain an invariant scalar, namely ${\mathcal S}$, and to perform the replacement $R\rightarrow R+f({\mathcal S})$ in the  E-H action (\ref{EH-action}), where $f$ is a vanishing function  at low energy densities, in order to recover  GR in this regime. 

\

On the other hand, 
if one wants that the modified Friedmann equation coming from this theory is a constraint and not a dynamical equation,  then one has to assume that
the scalar ${\mathcal S}$ which for the synchronous co-moving observers in the spatially flat Friedmann-Lema{\^\i}tre-Robertson-Walker (FLRW) spacetime, characterized by $ds^2=-dt^2+a^2(t)(dx_1^2+dx_2^2+dx_3^2)$,  is proportional to $H^2$,  where $H \equiv \dot{a}/a$, is the Hubble parameter of this universe (Here we note that the overhead ``dot'' is the usual one, that means it represents  the cosmic time differentiation). This property which guarantees that the  corresponding Hamiltonian  or the modified Friedmann equation is a constraint, actually means  that the quantity ${\mathcal S}$ couldn't be $R$, since in the flat FLRW spacetime 
for synchronous co-moving observers, the scalar curvature is derived to be 
$R=6(\dot{H}+2H^2)$.

\

In the same way one can show that  this scalar is not the square root of a linear combination of quadratic scalars  such as $R^2$, the Gauss-Bonnet (G-B) invariant or the Kretschmann scalar,
because when one one tries to remove the term $\dot{H}H^2$,  automatically $H^4$ is also  removed. 
However, if one considers the trace-free Ricci tensor
\begin{eqnarray}
 {\mathcal R}_{\mu\nu}\equiv R_{\mu\nu}-\frac{1}{4}Rg_{\mu\nu},
\end{eqnarray}
where $R_{\mu\nu}$ is the Ricci tensor, one could use the Carminati-McLenaghan invariants \cite{cm}
\begin{eqnarray}
 {\mathcal R}_2\equiv \frac{1}{4}{\mathcal R}_{\mu}^{\nu}{\mathcal R}_{\nu}^{\mu}, \quad \mbox{ and } \quad 
 {\mathcal R}_3\equiv -\frac{1}{8}{\mathcal R}_{\mu}^{\nu}{\mathcal R}_{\gamma}^{\mu}{\mathcal R}^{\gamma}_{\nu},
\end{eqnarray}
whose values for the synchronous co-moving observers 
in the flat FLRW spacetime are given by
${\mathcal R}_2=\frac{3}{4}\dot{H}^2$ and ${\mathcal R}_3=-\frac{3}{8}\dot{H}^3$. Thus, in such coordinates, one finds that, $\dot{H}=-2\frac{{\mathcal R}_3}{{\mathcal R}_2}$, and consequently, the scalar curvature takes the relation $-6H^2= -6\frac{{\mathcal R}_3}{{\mathcal R}_2}-\frac{R}{2}$. 
In a similar fashion, one also has $-6H^2=-(72 {\mathcal R}_3)^{1/3}-\frac{R}{2}$.
Following this, one can use the following scalars as invariants
\cite{amoros}
\begin{eqnarray}
 {\mathcal R}\equiv -6\frac{{\mathcal R}_3}{{\mathcal R}_2}-\frac{R}{2}, \quad \mbox{or} \quad \bar{\mathcal R}\equiv -(72 {\mathcal R}_3)^{1/3}-\frac{R}{2},
 \end{eqnarray}
which although seem to have an unusual structure but concerning with the cosmological perturbations, they deserve future investigations.
Effectively, at the background level the Friedmann equation will be a constraint relating the Hubble parameter and the energy density, because, due to our choice,  in the action, we only have the scale factor,  its derivative
and a divergence that do not affect the total dynamics. However,  dealing with the perturbations, since these scalars contain second derivatives of the metric as the Ricci curvature,   an action that contains the term 
$f({\mathcal R})$
or $f(\bar{\mathcal R})$, has the same degrees of freedom to that of $f(R)$ gravity. So, in principle, one may encounter with the Ostrogradski ghost problem which might be a disappointing  feature of the theory.

\

Nevertheless, as pointed out in \cite{yoshida}, one way to avoid the Ostrogradski instability is to introduce some contractions of the Weyl tensor and its products
for example $C_{\mu\nu\rho\sigma}C^{\mu\nu\rho\sigma}$ into the invariant scalars, since the Weyl's tensor vanishes in the spatially flat FLRW spacetime, and its inclusion does not affect the background dynamics. In addition, at the perturbative level, it can cure this problem. Moreover, in the same paper \cite{yoshida},
using the tensor $\nabla_{\mu}\nabla^{\nu}R$, the authors built up
a complicated gauge invariant scalar (see Section $3$ of \cite{yoshida})
that does not lead to the Ostrogradski ghost in the theory presented there. Thus, the use of this  scalar in our theory also deserves future investigation. 

\

A more usual way to obtain this scalar is to use of the Weitzenb\"ok connection, whose main scalar is the torsion ${\mathcal T}$, which for the flat FLRW geometry 
working in synchronous co-moving coordinates
is given by $-6H^2$. However,  as pointed out in \cite{lsb}, the main problem of this approach is that, a preferred orthonormal basis has to be chosen in the tangent bundle
because the theory is not local Lorentz invariant. Another  approach, based in the ADM formalism \cite{ADM} where it is assumed that the spacetime admits an slicing $\{\Sigma_t\}_{t\in \mathbb{R}}$,  is to consider the intrinsic curvature scalar
${\mathcal I}= K_{ij}K^{ij}-(Tr(K))^2$, where $Tr(K)=K_i^i$, is the trace of the extrinsic curvature tensor
\begin{eqnarray} 
 K_{ij}=
 g(\nabla_{e_i} {e_j}, {\bf n}),
 \end{eqnarray}
 where the Levi-Civita connection is denoted by $\nabla$;  ${\bf n}$ is the orthonormal vector field to $\Sigma_t~$; and
 $\{e_i\}_{i=1,2,3}$,  is a  basis in the tangent space of  $\Sigma_t$.   Once again, for the flat FLRW spacetime one can calculate that, ${\mathcal I}=-6H^2$,  but this approach is not gauge invariant in the sense that it depends on the  chosen slicing \cite{ha17}.
 
 \

 A totally different way to find bouncing backgrounds is via the so-called
  $F(R,T)$ gravity \cite{harko}, where $T$ denotes the trace of the stress tensor. In this theory, on the contrary to the usual proposals, apart form the gravitational sector,  it is  also the matter one which is modified. Dealing with the particular case $F(R,T)=R+ \frac{1}{M_{pl}^2} f(T)$, the modified Friedmann equation, for synchronous  co-moving observers in the flat FLRW spacetime becomes \cite{Shabani} 
 \begin{eqnarray}\label{xxx}
 3H^2=\frac{1}{M_{pl}^2}\left[(1+ f'(T))\rho+f'(T){\mathcal L}_{matt}-\frac{f(T)}{2}\right].
 \end{eqnarray}
 
The  main problem of this approach is that one has to express $\rho$ and ${\mathcal L}_{matt}$  as a function of $T$, and this is only possible in few cases, for example when the universe is filled with a perfect fluid or a scalar field mimicking the perfect fluid.
Another problem that follows is that, in this approach the conservation equation is different from the usual one,  which complicates very much the way to obtain bouncing backgrounds, and a few of them obtained seems to be 
unrealistic \cite{Shabani}. Moreover, the implementation of the cosmological perturbations is also unclear in this framework. Therefore,  this approach seems to deserve future investigations in order to clarify these unclear points.

 \
 
 Thus, {being motivated}
 to find a real  workable gauge invariant scalar quantity having  the desired property {that}, for the flat FLRW geometry, it becomes proportional to the square of the Hubble parameter only, first of all, we consider the stress tensor 
 (although  as we have stressed in the Introduction, since we want to modify the gravitational sector, it might  be better to consider the Ricci tensor), 
 \begin{eqnarray}
 T_{\mu}^{\nu}=\phi_{\mu}\phi^{\nu}-\left(\frac{1}{2}\phi_{\alpha}\phi^{\alpha}+V(\phi)   \right)\delta_{\mu}^{\nu},
\end{eqnarray}
where we use the notation $\nabla_{\mu}\phi\equiv \phi_{\mu}$. 
\

We can see that the normalized gradient field $\bar{\phi}^{\mu}\equiv \frac{\phi^{\mu}}{\sqrt{-\phi_{\alpha}\phi^{\alpha}}}$ is an eigenvector of the stress tensor, and basically we will use it
as the gradient of the mimetic field introduced in \cite{mukhanov}. For synchronous co-moving  observers in the spatially flat FLRW spacetime one has the form $\bar{\phi}=(\pm 1, 0,0, 0)$, and 
we can use  the following two  elementary Lagrangians of Degenerate  Higher Order Scalar Theories (DHOST), namely 
$( \nabla_{\mu}\bar\phi^{\mu})^2$ and $ \nabla^{\nu}\bar\phi^{\mu}\nabla_{\nu}\bar\phi_{\mu}$ (see Section $2$A of \cite{langlois}).
Now,  in an analogous way  to the extrinsic curvature scalar definition, we introduce the scalar $\Phi\equiv   \nabla^{\nu}\bar\phi^{\mu}\nabla_{\nu}\bar\phi_{\mu}    -( \nabla_{\mu}\bar\phi^{\mu})^2$, which in the flat FLRW
 spacetime,
 for synchronous co-moving observers, leads to $-6H^2$.  This is the invariant we will use in this note to build bouncing backgrounds, and as one can easily realize, is  essentially the same used 
 in \cite{langlois}. 
 Although one can simply use the divergence of $\bar{\phi}^{\mu}$ to obtain the scalar $\chi\equiv-\nabla_{\mu}\bar{\phi}^{\mu}$ which for the flat FLRW geometry becomes,
 $3H$.

 \
 
 To obtain the dynamical equations, in the flat FLRW background, we work in the coordinates with the line element
 $ds^2=-N^2dt^2+a^2(t)(dx_1^2+dx_2^2+dx_3^2)$,  where $N(t)$ is the lapse function. The  modified action we consider is 
 \begin{equation}\label{action}
  S_{f}=\int N{\mathcal V}\left(\frac{M_{pl}^2}{2}\left(R+f(\Phi)\right)+{\mathcal L}_{matt}\right) dt,
 \end{equation}
where  ${\mathcal V}\equiv a^3$ is the volume and the  matter Lagrangian is given by
 \begin{eqnarray}
  {\mathcal L}_{matt}
  =\frac{\dot{\phi}^2}{2N^2}
  -V(\phi).
  \end{eqnarray}
 
 In this coordinates one has $\Phi=-\frac{6H^2}{N^2}$ and $R=6\left( \frac{1}{aN}\frac{d}{dt}\left(\frac{\dot{a}}{N}  \right)     +\frac{H^2}{N^2}\right)$. Now, since
 \begin{eqnarray}
 {\mathcal V} N\frac{1}{aN}\frac{d}{dt}\left(\frac{\dot{a}}{N}  \right) = \frac{d}{dt}\left(a^2\frac{\dot a}{N}\right)-2{\mathcal V}N\frac{H^2}{N^2},
   \end{eqnarray}
 the action  (\ref{action})  is equivalent to the following one
  \begin{equation}\label{action-modified}
  \bar{S}_{f}=\int N{\mathcal V}\left(\frac{M_{pl}^2}{2}\left(\Phi+f(\Phi)\right)+{\mathcal L}_{matt}\right) dt.
 \end{equation}

 The variation of (\ref{action-modified}) with respect $N$ leads to the Hamiltonian constraint, which is equivalent to the modified Friedmann equation, and the variation of (\ref{action-modified}) with respect the volume leads to the dynamical equation or the Raychaudhuri equation. So, doing it, and taking $N=1$, to work in the synchronous gauge, we obtain
\begin{eqnarray}\label{eqn-Friedmann}
 -2\Phi f_{\Phi} +f-  \Phi =\frac{2\rho}{M_{pl}^2},
  \end{eqnarray}
  \begin{eqnarray}
  \dot{\Phi}\left( f_{\Phi}-2\Phi f_{\Phi\Phi}+1      \right)=\frac{6H}{M_{pl}^2}(\rho+P),
    \end{eqnarray}
  where $f_{\Phi}$ is the partial derivative of $f$ with respect to $\Phi$, and the energy density is $\rho=\frac{\dot{\phi}^2}{2}+V(\phi).$

 \begin{remark}
 In the same way, using the invariant $\chi$, the dynamical equations will be
 \begin{eqnarray}
 -\chi f_{\chi} +f-  \frac{\chi^2}{3} =\frac{\rho}{M_{pl}^2},
  \end{eqnarray}
  \begin{eqnarray}
  \dot{\chi}\left( 1-\frac{3}{2}\Phi f_{\Phi\Phi}      \right)=-\frac{3}{2M_{pl}^2}(\rho+P).
    \end{eqnarray}
 \end{remark}

 \
 
 Therefore, for any curve, $\rho=\frac{M_{pl}^2}{2}g(\Phi)$, described in the plane $(\Phi,\rho)$, the differential equation (\ref{eqn-Friedmann}) reduces to 
 \begin{eqnarray}
  -2\Phi f_{\Phi} +f-  \Phi =g(\Phi), \end{eqnarray}
which after one-time integration gives the corresponding $f$ with the following solution
 \begin{eqnarray} \label{fflqc}
 f(\Phi)=
 -\frac{\sqrt{-\Phi}}{2}\int
  \left( \frac{g(\Phi)}{\Phi\sqrt{-\Phi}} \right) d{\Phi}-\Phi,
 \end{eqnarray}
 and the same process also be done using the scalar $\chi$ instead of $\Phi$.

\section{The simplest bouncing scenario}
\label{sec-bouncing}

 Dealing with the flat FLRW geometry   and using synchronous co-moving coordinates, we 
 redefine the Hubble parameter as $\bar{H}=k M_{pl}^3H$ where $k$ is a dimensionless constant, in order that $\bar H$ has the  units of energy density.
   In the plane $(\bar{H},\rho)$, the simplest closed curve is a circle,  and taking into account that the energy density has to be positive and the Hubble parameter must be zero when the energy density vanishes, we must choose  a circle
 centered at  $(0,\bar{\rho})$ with radius 
 $\bar{\rho}$,  being $\bar{\rho}$
  a constant with units of energy density, that is, 
  $\bar{H}^2+(\rho-\bar{\rho})^2=\bar{\rho}^2$, which in the plane $(H,\rho)$  will depict 
 the  ellipse 
 \begin{eqnarray}\label{friedmann}
  H^2=\frac{\rho}{k^2M_{pl}^6}\left( 2\bar{\rho}-{\rho} \right).
 \end{eqnarray}

Moreover, at low energy densities one has to recover the Friedmann equation $H^2=\frac{\rho}{3M_{pl}^2}$, then we must  impose $6\bar\rho=k^2M_{pl}^4$, and the equation (\ref{friedmann}) becomes
 \begin{eqnarray}\label{lqc}
  H^2=\frac{\rho}{3M_{pl}^2}\left( 1-\frac{\rho}{2\bar\rho} \right),
 \end{eqnarray}
which can be looked as the
holonomy corrected Friedmann equation in LQC with 
the replacement of $\bar\rho$ by $\rho_c/2$.

\

On the other hand,   
 since the equation (\ref{lqc}) could be written as a bi-valued function
 \begin{eqnarray}
 \rho=\frac{\rho_c}{2}\left(1\pm\sqrt{1+\frac{2\Phi M_{pl}^2}{\rho_c}}\right),
 \end{eqnarray}
 where the sign ``$-$''  (respectively ``$+$'') correspond to the lower  (respectively upper) branch of the ellipse, thus, it is clear that in order 
  to depict this constraint we need a bi-valued function $f$,  which could easily be obtained after integrating the equation (\ref{fflqc}) having \cite{helling, ds09, haro12}:
 \begin{eqnarray}
 f(\Phi)=\frac{\rho_c}{M_{pl}^2}\left(1- \sqrt{1-s^2}- s\arcsin(s)  \right)-\Phi,
 \end{eqnarray}
 where
 $s\equiv \sqrt{-\frac{2\Phi M_{pl}^2}{\rho_c}},$
 and the sign of the square root has been chosen to be positive (respectively negative) in the lower  (respectively upper) branch  and
$ \arcsin(s) \equiv \int_0^s \frac{1}{\sqrt{1-{\bar s}^2}}  {d\bar{s}}$ in the lower branch whereas  $\arcsin(s) \equiv \int_0^s \frac{1}{\sqrt{1-{\bar s}^2}}  {d\bar{s}}+\pi$, in the upper one, with the same criteria for the sign of the square root.

  \
  
In general,
  at low energy densities a viable theory must coincide with GR which means that in case of a spatially flat FLRW geometry,  the modified  Friedmann equation 
should be $H^2=\frac{\rho}{3M_{pl}^2}$. An immediate consequence of this relation exibit that for $\rho\cong 0$, one should have $H \cong 0$.  This means that, at the bouncing backgrounds the energy density has to be a multi-valued function because at the same time it has to be zero and different from it (at the bounce) when the Hubble parameter vanishes,  and consequently, to reproduce these bouncing backgrounds multi-valued  $f$-functions will be needed.

 \
 
 Finally, once the background is obtained, recall that when one deals with a scalar field there are infinitely many backgrounds, obtained from the equation
 \begin{eqnarray}\label{Conservation}
 \ddot{\phi}+3H(\rho)\dot{\phi}+V_{\phi}=0,
 \end{eqnarray}
 one has to deal with the cosmological perturbations.  
 This topic has been studied for  modified teleparallel gravity in \cite{cai} and applied to the particular case of a function $f$ mimicking the holonomy corrected LQC background in 
 \cite{ha14a} obtaining equations for scalar and tensor perturbations that differ from the ones of LQC \cite{grain,caitelleau} in the velocity of sound,  using the ADM formalism 
 applied to the $f$-theory leading to the same background as LQC it has been obtained in \cite{ha17} that for scalar perturbations the perturbation equation is the same as in LQC, and for tensor perturbations they only differ in the velocity of sound. 
Recently, in mimetic gravity it has been found that at the background level such $f$-theory can mimick LQC \cite{hap} with a suitable Lagrange's multiplier. In addition to that, when the scalar perturbations are concerned in such framework, 
this modified mimetic 
$f$-theory returns similar equations to that of LQC but for the tensor perturbations, the equations do not differ the usual equations obtained in GR. 
 
 \

Studying cosmological perturbations, for the moment,  in our approach and  dealing with the  scalar $\chi$ instead of $\Phi$,
we have obtained the general equation \cite{ha}
\begin{eqnarray}\label{aaa}
G_{\mu\nu}=T_{\mu\nu}+\tilde{T}_{\mu\nu}
\end{eqnarray}
where $G_{\mu\nu}=R_{\mu\nu}-\frac{1}{2}g_{\mu\nu}R$, is the Einstein tensor and
\begin{eqnarray}\label{bbb}
\tilde{T}_{\mu\nu}
\equiv
\left(f-\chi f_{\chi}+\bar{\phi}^{\alpha}\chi_{,\alpha}f_{\chi\chi} 
\right)g_{\mu\nu} \nonumber\\
-f_{\chi\chi}(\bar{\phi}_{\nu}\chi_{,\mu}+\bar{\phi}_{\mu}\chi_{,\nu}
+\bar{\phi}^{\alpha}\chi_{,\alpha}\bar{\phi}_{\mu}\bar{\phi}_{\nu}).
\end{eqnarray}
The conservation equation takes the form 
\begin{eqnarray}\label{ccc}
-\Box \phi+ V_{\phi}+\nonumber\\ \nabla_{\mu}\left(  \frac{1}{\sqrt{-\phi^{,\alpha}\phi_{,\alpha} }} \left( \partial^{\mu}f_{\chi}+
\bar{\phi}^{\mu}\bar{\phi}_{\alpha}\partial^{\alpha}f_{\chi}   \right) \right)=0, 
\end{eqnarray}
which for the flat FLRW spacetime leads to the usual conservation equation (\ref{Conservation}).

\

Now, considering the tensor perturbations, since the field $\phi$ does not affect them,  one has the same equations as in GR. On the contrary, for scalar perturbations when
one uses the  longitudinal gauge, 
from the equations (\ref{aaa}) and (\ref{bbb}) one can see that for the Newtonian potential, namely $\Psi$, it has the time derivatives up to order $2$ (i.e., we do not encounter with the Ostrogradski's instability) because 
$\chi=3H-3(\dot{\Psi}+H\Psi)-\frac{1}{a^2\dot{{\phi}}_0}\Delta\delta\phi$, where 
$\phi_0$ denotes the homogenous part of the field $\phi$ and
$\delta\phi$ is the perturbed part. In fact, after a cumbersome calculation one 
finds 
\begin{eqnarray} \label{ddd}
{\ddot{\Psi}}-\Omega\frac{\Delta\Psi}{a^2}+\left(H-2\frac{\ddot{\phi}_0}{\dot{\phi}_0}
-\frac{\dot{\Omega}}{\Omega}\right){\dot{\Psi}}+ 
\nonumber\\
\left(2\left(\dot{H}-H\frac{\ddot{\phi}_0}{\dot{\phi}_0}\right)
-H\frac{\dot{\Omega}}{\Omega} \right){\Psi}
=\Omega
\frac{{\dot{{\phi}}_0}^2}{2}\partial_t\left(\frac{{f}_{\chi\chi}\Delta\delta\phi}{a^2{\dot{{\phi}}}_0^3}   \right),
\end{eqnarray}
where $\Omega=\frac{1}{1-\frac{3}{2}f_{\chi\chi}}$. 
In the same way the dynamical equation for $\delta \phi$ can be obtained from (\ref{ccc}), and one can also check that the perturbed part of the scalar field only contains the time derivatives up to order $2$, and thus, Ostrogradski's instability does not appear. We note that the disappearance of the Ostrogradski instability is not necessary for all orders of the perturbations. For the higher-order of the perturbations, it is expected to have such instability. Thus, we stress that the vanishing of the Ostrogradski instability is true only for the linear level of the perturbations not for all orders of the perturbations for this particular model.

Moreover, from (\ref{ddd})
 as one can see that, although the 
ghost instability does not appear because the second derivative of the Newtonian potential $\Psi$ has the right sign, but the
gradient instabilities may appear due to the factor $\Omega$ multiplied by the Laplacian of $\Psi$.  Note that when one deals with the $f$-theory that mimicks LQC, this equation 
is not exactly the same as obtained in LQC, because the right hand side term of eqn. (\ref{ddd}) does not vanish. But it has the same features, because in that case 
$\Omega=1-\frac{2\rho}{\rho_c}$, being $\rho_c$ the critical energy density defined at the beginning of the section. So, for energies between $\frac{\rho_c}{2}$ and $\rho_c$, i.e., in the upper branch of the ellipse,  the gradient instability appears, but as it has been discussed in some bouncing scenarios, namely, the matter bouncing \cite{wilson-ewing} or matter-ekpyrotic bouncing cosmologies \cite{haa}, the pivot scale leaves the Hubble radius in the lower branch, henceforth, this instability does not affect the observable modes.

Of course, this new approach based on the scalar $\Phi$ or $\chi$ deserves future investigations which include the comparison with the perturbative LQC equations, in particular.

\section{Conclusions}
\label{sec-discuss}

 In this work we focus on the construction of viable bouncing cosmologies in the context of modified gravity theories. 
In general, bouncing backgrounds in modified gravitational theories are very  difficult to obtain and the generated models are often criticized for some sensitive issues. The simplest bouncing scenarios obtained in the modified  teleparallel theories and in the  modified gravity theories based
on the ADM formalism  suffer with a similar problem. The former one is not  Lorentz invariant while the last one is not gauge invariant. That means, although one could be able to construct bouncing cosmologies in those modified gravity theories, but the lack of Lorentz and gauge invariances in the aforementioned modified theories raise a question mark on the resulting scenarios. It is clear that a modified version of any gravitational theory without such foregoing problems is surely interesting for further investigations and could open some new possibilities.

 \

Thus, in the present work, based on the idea that the matter component of the universe could be used to build up scalars, we have tried to obtain a Lorentz and gauge invariant theory with the use of quadratic combinations of the 
covariant derivative of the time-like unitary eigenvalues of the stress tensor.
For the synchronous co-moving observers in the flat FLRW spacetime this scalar reduces to $-6H^2$, which ensures that the corresponding modified Friedmann equation imposes a constraint between $H$ and $\rho$ of the FLRW universe,  and consequently, this allows us  to obtain the bouncing backgrounds when this modified Friedmann equation depicts a closed curve in the phase space $(H,\rho)$.
Finally, we show that in order to have a bounce, the use of a multivalued function $f$  is mandatory, and the simplest one is the one that leads to the holonomy corrected Friedmann equation in LQC. 
Working in this context when an invariant scalar plays the key role, we must be cautious about the stability of the model at its perturbative level since the scalar invariants include the  second or higher order time derivatives of the  metric, and hence the Ostrogradski instability could be a natural outcome for such models \cite{yoshida}. In the present work, dealing with the perturbations at the linear level, 
our theory leads, for tensor perturbations, to the same equations as in GR
and for the scalar perturbations in the longitudinal gauge, the highest order derivative of the Newtonian potential and the perturbed part of the scalar field is two. As a consequence, we do not find any Ostrogradski instability while the gradient instabilities may still appear in such theories. It is important to stress that in the present work, the disappearance of the Ostrogradski instability is valid only at the linear level of the perturbations while for all higher-orders of the perturbations, it 
is necessarily not true. In other words, the Ostrogradski instability might not be completely removed from the present context. 

\

Thus, although much further investigations are necessary in order, the way presented in the current work might be considered as an initiation towards the viable constructions of bouncing cosmologies in modified gravity theories keeping the Ostrogradski instabilities and other instabilities as well in the picture.

\section*{Acknowledgments}
We thank the referee for raising some important comments. We also thank Professor Jaume Amor\'os for helpful discussions and Llibert Arest\'e Sal\'o for helping us performing the calculations in the perturbative case.  
This investigation has been supported in part by MINECO (Spain), project MTM2014-52402-C3-1-P.

\end{document}